\documentclass[preprint]{aastex}
\usepackage{mkfig}

\begin{document}

\title{\bf Deuterium Toward WD~1634-573: Results from the Far Ultraviolet
  Spectroscopic Explorer (FUSE) Mission\altaffilmark{1}}

\author{B. E. Wood,\altaffilmark{2} J. L. Linsky,\altaffilmark{2}
  G. H\'{e}brard,\altaffilmark{3} A. Vidal-Madjar,\altaffilmark{3}
  M. Lemoine,\altaffilmark{3} H. W. Moos,\altaffilmark{4}
  K. R. Sembach,\altaffilmark{4} and E. B. Jenkins\altaffilmark{5}}

\altaffiltext{1}{Based on observations made with the NASA-CNES-CSA Far
  Ultraviolet Spectroscopic Explorer.  FUSE is operated for NASA by the
  Johns Hopkins University under NASA contract NAS5-32985.}
\altaffiltext{2}{JILA, University of Colorado and NIST, Boulder, CO
  80309-0440; woodb@casa.colorado.edu.}
\altaffiltext{3}{Institut d'Astrophysique de Paris, CNRS, 98 bis Boulevard
  Arago, F-75014 Paris, France.}
\altaffiltext{4}{Department of Physics and Astronomy, Johns Hopkins
  University, 3400 North Charles Street, Baltimore, MD 21218.}
\altaffiltext{5}{Princeton University Observatory, Princeton, NJ 08544.}

\begin{abstract}

     We use {\em Far Ultraviolet Spectrocopic Explorer} (FUSE) observations
to study interstellar absorption along the line of sight to the white dwarf
WD~1634-573 ($d=37.1\pm 2.6$~pc).  Combining our measurement of D~I with a
measurement of H~I from {\em Extreme Ultraviolet Explorer} data, we find a
D/H ratio toward WD~1634-573 of ${\rm D/H}=(1.6\pm 0.5)\times 10^{-5}$.
In contrast, multiplying our measurements of
${\rm D~I/O~I}=0.035\pm 0.006$ and ${\rm D~I/N~I}=0.27\pm 0.05$ with
published mean Galactic ISM gas phase O/H and N/H ratios yields
${\rm D/H}_{O}=(1.2\pm 0.2)\times 10^{-5}$ and
${\rm D/H}_{N}=(2.0\pm 0.4)\times 10^{-5}$, respectively.  Note that all
uncertainties quoted above are 2$\sigma$.
The inconsistency between ${\rm D/H}_{O}$ and ${\rm D/H}_{N}$ suggests that
either the O~I/H~I and/or the N~I/H~I ratio toward WD~1634-573 must be
different from the previously measured average ISM O/H and N/H values.  The
computation of ${\rm D/H}_{N}$ from D~I/N~I is more suspect, since the
relative N and H ionization states could conceivably vary within the LISM,
while the O and H ionization states will be more tightly coupled by charge
exchange.

\end{abstract}

\keywords{white dwarfs --- stars: individual (WD~1634-573) --- ISM:
  abundances --- ultraviolet: ISM}

\section{INTRODUCTION}

     The deuterium-to-hydrogen (D/H) ratio is of central importance to
many areas of astrophysics.  In the standard Big Bang cosmology, the
primordial D/H value places strong constraints on the total baryon content
in the universe.  Since the deuterium abundance is believed to
decrease with time due to the destruction of deuterium in stellar interiors,
measurements of the local D/H value in our galaxy provide a lower limit for
the primordial D/H value, while measurements of D/H in the Lyman-$\alpha$
forest provide an estimate that may be closer to the true primordial D/H
ratio.  Comparing the local and Lyman-$\alpha$ forest values provides a
measure of how much stellar processing of material has occurred during the
lifetime of the Milky Way Galaxy.  Measurements of D/H in different
locations in the Galaxy can provide critical tests of Galactic chemical
evolution models.

     The {\em Copernicus} satellite provided the first accurate D/H
measurements within the Galactic ISM.  \citet{jbr73} measured
${\rm D/H}=(1.4\pm 0.2)\times 10^{-5}$ toward $\beta$~Cen, and
\citet{dgy76} quoted an average value of
${\rm D/H}=(1.8\pm 0.4)\times 10^{-5}$ for several lines of sight observed
by {\em Copernicus}.  The Local Interstellar Cloud (LIC) within which the
Sun resides appears to have ${\rm D/H}=(1.5\pm 0.1)\times 10^{-5}$ based on
{\em Hubble Space Telescope} (HST) measurements along many lines of sight
through the LIC \citep{jll98}, with no evidence for any significant
variation within this small cloud \citep[$\sim 7$ pc across based on a
compilation of LIC measurements;][]{sr00}.  Estimating the primordial
value from the Lyman-$\alpha$ forest has proved to be more difficult,
although these measurements do suggest a higher D/H in the Lyman-$\alpha$
forest compared with the LIC, as one would expect
\citep[e.g.,][]{jkw97,sb98,dt99,dk00}.  For a more thorough review of these
intergalactic D/H measurements and how they relate to Galactic D/H,
see \citet{hwm01}.

     However, the LIC D/H value quoted above is also problematic, as it
measures only very local interstellar gas.  Values of D/H elsewhere in the
Galaxy could in principle be much different, since material in different
parts of the Galaxy has undergone varying amounts of stellar processing.
Evidence for variations of D/H within the LISM have in fact been found for
longer lines of sight sampling many clouds, based on data from HST and the
IMAPS instrument \citep{avm98,ebj99,gs00,ml01}.  Measuring the variation of
D/H within the Galaxy is important both for
determining the degree of mixing of ISM material within the Milky Way, and
for determining a better estimate for the mean present day Galactic D/H
value.  Unfortunately, HST is not able to obtain measurements of D/H for long
lines of sight through the Galaxy, because for column densities greater than
$5\times 10^{18}$ cm$^{-2}$ the H~I Ly$\alpha$ absorption becomes so broad
that it completely obscures the D I absorption.

     Measuring deuterium at different locations throughout our Galaxy
is one of the primary missions of the {\em Far Ultraviolet Spectroscopic
Explorer} (FUSE).  Unlike HST, FUSE can observe the higher lines of the
H~I Lyman series, which allows deuterium to be observed for higher ISM
column densities.  This paper is one of a series reporting the first
deuterium measurements made with FUSE \citep{sdf01,gh01,jwk01,nl01,ml01,gs01},
including a summary paper of these
initial results \citep{hwm01}.  The line of sight analyzed here is
that toward the white dwarf WD~1634-573 (=HD~149499B).  Table~1 lists
properties of our target star, which is a very metal poor DO white dwarf
unlikely to have photospheric metal lines contaminating the ISM absorption
lines in which we are interested \citep{sd99}.  There is a K0~V companion
star (HD~149499A) that will be in the FUSE aperture for our observations.
However, even though this star is believed to be quite active \citep{sj97},
emission from this star will not be detectable against the
very bright continuum of the hot white dwarf, based on extrapolations of
H~I Lyman line fluxes observed for other active single K dwarfs.

     With a path length of only $37.1\pm 2.6$~pc \citep{macp97},
the WD~1634-573 line of sight does not exceed the distance scale of
deuterium measurements accessible to HST.  However, the H~I column density
estimated from {\em Extreme Ultraviolet Explorer} (EUVE) data is
$N_{\rm H}=(7\pm 2)\times 10^{18}$ cm$^{-2}$ \citep{rn96},
too high for deuterium to be detectable in Ly$\alpha$, so this line of
sight {\em does} extend deuterium measurements to higher columns than are
measurable with HST.

\section{FUSE OBSERVATIONS}

     The DO white dwarf WD~1634-573 has been observed by FUSE three times
through the low resolution (LWRS) aperture and twice through the medium
resolution (MDRS) aperture.  These observations are listed in Table~2.
Collectively, the LWRS data set consists of
28 separate exposures with a total exposure time of 19,406 s, while the
MDRS data set consists of 37 separate exposures totaling 18,225 s.

     In order to fully cover its 905--1187~\AA\ spectral range, FUSE has a
multi-channel design --- two channels (LiF1 and LiF2) use Al+LiF coatings,
two channels (SiC1 and SiC2) use SiC coatings, and there are two different
detectors (A and B).  For a full description of the instrument, see
\citet{hwm00}.  With this design FUSE acquires spectra in eight segments
covering different, overlapping wavelength ranges.  For both of the LWRS and
MDRS data sets, the individual exposures processed in the standard FUSE
pipeline (version 1.6.9) are cross-correlated and coadded to create a
single spectrum for each segment.  We decided not to coadd the individual
segments to ensure that such an operation would not degrade the spectral
resolution.  Instead, a spectrum covering the full FUSE spectral range is
spliced together from several different segments.  To be more precise, the
spectrum combines the following segments:  SiC1B (910--925~\AA), SiC2A
(925--995~\AA), LiF1A (995--1075~\AA), SiC2B (1075--1095~\AA), and LiF2A
(1095--1180~\AA).

     We keep the LWRS and MDRS spectra separate, since the spectral
resolution of the two may be slightly different.  This also gives us two
separate spectra that we can analyze independently to see if we obtain the
same answer from both data sets (see below).  One exception is the
H~I Ly$\beta$ line at 1025~\AA.  For the LWRS spectrum, there is substantial
contamination from airglow emission in this line, which is not the
case in the MDRS data thanks to the narrower aperture.  Thus, we import
the MDRS Ly$\beta$ line into the LWRS spectrum to allow the analysis of the
LWRS data to also consider this important line.  Airglow is also a potential
contaminant for many of the O~I and N~I lines \citep{pdf01}, but
the continuum fluxes of WD~1634-573 are high enough to overwhelm the airglow
and we see no difference in the LWRS and MDRS spectra that would suggest a
problem.

     Since there is no wavelength calibration lamp onboard FUSE, the
wavelength scale produced by the FUSE data reduction pipeline is very
uncertain.  Thus, we identify many ISM absorption lines throughout the LWRS
and MDRS spectra and measure their centroids.  For each segment, the average
shift between the centroids and rest wavelengths of lines within that
segment is used to correct the wavelength scale and place it in the rest
frame of the interstellar absorption.

     Figure~1 shows the resulting reduced spectrum for the LWRS data.
Coadding the separate exposures of these segments helps to suppress fixed
pattern noise somewhat.  However, the spectrum in Figure~1 shows that
significant periodic fixed pattern noise remains in many locations, in
particular the moir\'{e} pattern mentioned by \citet{djs00} that
produces noise with a period of about nine pixels.  The effective
signal-to-noise of our spectrum is in many locations limited by the moir\'{e}
ripples rather than actual photon noise.

\section{ANALYSIS}

\subsection{GHRS Data}

     Obtaining the most accurate measurements for column densities requires
that the number of ISM absorption components along the line of sight be
known with some precision.  Unfortunately, FUSE does not have sufficient
spectral resolution to resolve the ISM velocity structure unless the
components are very widely separated, and there are also no high resolution
HST or ground based data to provide this information.  There are, however, a
few moderate resolution HST spectra of WD~1634-573 taken on 1996 September 3
with the G160M grating of the Goddard High Resolution Spectrograph (GHRS)
instrument formerly on board HST, and these spectra contain a few ISM lines
\citep{sd99}.  In Figure~2, we plot the ISM absorption lines of
S~II $\lambda$1259.519 and Si~II $\lambda$1260.422 on a heliocentric velocity
scale.  We have fitted the S~II line with a single absorption component, with
a correction for instrumental broadening assuming a Gaussian instrumental
profile with a width of 4.4 pixels \citep{rlg94}.

     The S~II line is centered at $-19.6$ km~s$^{-1}$, which is very
close to the velocity predicted by the LIC vector \citep{rl95},
$-18.9$ km~s$^{-1}$.  Actually, with Galactic coordinates of
$l=329.9^{\circ}$ and $b=-7.0^{\circ}$, the WD~1634-573 line of sight is
more likely to sample the so-called ``G cloud,'' a proposed very nearby
cloud in the Galactic Center direction very similar to the LIC in most
respects, but which has a slightly different flow vector \citep{rl92,bew00}.
The G cloud vector predicts a
velocity of $-21.2$ km~s$^{-1}$, a bit farther from the measured value of
$-19.6$ km~s$^{-1}$ than the LIC value, but still close enough to be
consistent considering this is moderate rather than high resolution data.

     However, when the stronger, opaque Si~II $\lambda$1259.519 line is
fitted with a single absorption component, the measured absorption centroid
is $-15.2$ km~s$^{-1}$, a 4.4 km~s$^{-1}$ discrepancy with S~II.  An
analogous effect is detectable in the FUSE data, where optically thick lines
also appear to be redshifted relative to optically thin lines nearby in the
spectrum.  This fact strongly suggests the presence of at least one extra
velocity component along the line of sight, which is redshifted relative to
the primary absorption component.  Thus, in Figure~2 the Si~II line is
fitted with two components, one with a velocity fixed at the velocity
suggested by the S~II line, and the other located redward of the main
component by 17 km~s$^{-1}$ to account for the extra absorption on that
side of the line.  The Doppler broadening parameters are forced to be the
same for both components to reduce the number of free parameters.

     Unfortunately, at the resolution of the GHRS G160M grating
($R\approx 20,000$) the ISM lines are not resolved any more than they are for
FUSE, so the true velocity structure of the WD~1634-573 line of sight is
still unknown.  The fit to the Si~II line in Figure~2 suggests at least one
additional component redward of the primary component, but the fit is not
quantitatively useful for accurately estimating component velocity
separations or column density ratios.

\subsection{Fitting the FUSE Data}

     Figure~1 provides a broad overview of the WD~1634-573 spectrum using
the LWRS data.  Figure~3 shows closeups of many of the ISM absorption lines
using MDRS data.  The WD~1634-573 spectral regions shown in Figure~1 contain
all of the absorption lines that we use to derive interstellar column
densities for this line of sight for various elements.  At least 21 H~I Lyman
lines are observable, from Ly$\beta$ at 1025~\AA\ down to the numerous H~I
lines clustered near the Lyman limit at 912~\AA.  The six strongest Lyman
lines are plotted in Figure~4 on a velocity scale centered on the H~I
absorption, showing that deuterium
(D~I) absorption is detectable at about $-82$ km~s$^{-1}$ in at least the
strongest four lines (Ly$\beta$ through Ly$\epsilon$).

     Many O~I and N~I lines of varying strengths are also apparent in the
FUSE spectrum (see Fig.~1), meaning FUSE data should provide accurate
measurements of O~I and N~I column densities.  Other atomic species
with at least one detectable ISM line are N~II, P~II, C~II, C~III, Ar~I,
Fe~II, and perhaps Si~II.  The Si~II $\lambda$989 line may be
blended with a N~III line at that location \citep{ebj00} and its
strength seems to be inconsistent with the nondetection of
Si~II $\lambda$1020, so that identification is listed with a question mark
in Figure~1.

     As described by \citet{hwm01}, the first FUSE D/H analyses
published in this series of articles are all the result of at least two
independent analyses, and the study of WD~1634-573 presented here is no
exception.  This approach of using independent analyses with significantly
different methods is the best way to estimate the magnitude of systematic
errors, which are very difficult to quantify otherwise.  The final column
densities that we derive are compromise values consistent with the results
of both of these independent analyses.  Several of us (G.\ H., M.\ L., and
A.\ V.-M.) analyzed the WD~1634-573 data in a manner very similar to the
analysis of the WD~2211-495 line of sight described by \citet{gh01},
and we refer the reader to that paper for a thorough description.
We now describe the other independent analysis in far more detail.

     In order to analyze the absorption lines, we first estimate
the stellar continuum overlying the absorption.  This is done with a series
of polynomial fits covering the entire FUSE spectrum.  The result is shown
in Figure~1.  Estimating the continuum above the Ly$\beta$ $\lambda$1025
absorption requires a little more work because of the significant amount of
absorption in the line wings.  This wing absorption indicates that the
Lorentzian wings of the opacity profile are starting to become important,
which in turn signifies that the line is no longer completely in the flat
part of the curve of growth.  Our initial guess for the background continuum
assumed no wing absorption.  After the initial fit was performed (as
described below), the fit to the Ly$\beta$ spectral region was
poor because of the presence of wing absorption, so the residuals of that
fit were used to derive a higher continuum that would improve the quality of
the fit.  Another iteration leads to the Ly$\beta$ continuum shown in
Figures 1 and 4.

     Our fits to the absorption lines are global fits, where all the lines
are fitted simultaneously.  We use \citet{dcm91} as our primary source for
the necessary atomic data and \citet{dav94} for some of
the higher H~I Lyman lines.  Note that we use the \citet{dcm91} line list
available on the WWW
(\verb$http://www.hia.nrc.ca/staff/dcm/atomic_data.html$),
for which many of the N~I oscillator strengths have been updated from their
original published values.  All clearly detected
lines are identified in Figure~1 and also listed in Table~3 along with rest
wavelengths and oscillator strengths.  Table~3 lists equivalent widths
for the lines, except for cases where blending is too severe for an
accurate measurement.  The equivalent widths and quoted 2$\sigma$
uncertainties are based on measurements of both the LWRS and MDRS spectra,
assuming several different background continua estimations.

     However, in our fits not only are the
detected lines included, but we also consider many undetected lines of atomic
species with at least one detected line.  These nondetections can provide
useful upper limit constraints for our fitting routine.  There are two O~I
lines listed by \citet{dcm91} that we do {\em not} include in our fits
because their listed f values suggest line strengths that are clearly
inconsistent with the observations, based on comparison with the other O~I
lines seen in the spectrum.  These are the O~I $\lambda$972.1 and
$\lambda$1026.5 lines.  The oscillator strengths listed for these two
lines must be incorrect.

     The primary free parameters of the global fit are the column densities
of the various atomic species (H~I, D~I, O~I, N~I, etc.).  For all detected
and unblended lines, we allow the centroid of the line to vary as a free
parameter of the fit to account for inaccuracies in the wavelength
calibration.  The temperature ($T$) and nonthermal velocity
($\xi$) of the interstellar gas are also free parameters.  The Doppler
broadening parameters ($b$) of the various ISM lines are computed from $T$
and $\xi$ using the equation $b^{2}=0.0165T/A + \xi^{2}$, where $A$ is the
atomic weight of the atomic species associated with each line.  Note,
however, that because the lines are not resolved and because we cannot
accurately account for the multiple velocity components present in the lines
(see \S 3a), the validity of the $T$ and $\xi$ values derived
in our fits is questionable.

     In Figures~1 and 3, a single-component global fit to the absorption
lines is shown both before (dotted line) and after (thick solid line)
convolution with the instrumental profile.  Figure~1 shows a fit to the LWRS
data and Figure~3 shows a fit to the MDRS data.  Unfortunately, the
instrumental profile, or line spread function (LSF), is not currently well
known for FUSE, and may in any case conceivably vary from one observation to
the next depending on the accuracy of the pointing for that particular
observation and on how accurately the individual exposures are
cross-correlated and coadded \citep{jwk01}.  Thus, the LSF is a free
parameter of our fit.  We assume a two-Gaussian representation for the LSF,
which we assume to be applicable over the entire FUSE spectral region.  This
is just an approximation, however, as the LSF undoubtedly varies with
wavelength.

     The fits in Figures~1 and 3 are single-component fits
performed as described above.  In order to decide on the final column
density values and their uncertainties, however, we experiment with many
other fits.  Analyzing the LWRS and MDRS spectra
independently provides some sense of the systematic
uncertainties involved in the fitting process.  We also experiment with
different estimates for the continuum placement.  Particularly for
the D~I lines, we also try fits without considering the H~I
lines at all.  This allows us to more simply extrapolate a continuum just
over the D~I lines themselves rather than making the more uncertain
extrapolation over both the D~I and H~I lines, which are partially blended
for Ly$\beta$ and Ly$\gamma$ (see Fig.~4).  There is significant potential
for systematic error with regards to D~I since all the D~I lines are just
blueward of H~I lines, which makes continuum placement tricky.  The fact that
photospheric He~II and H~I absorption are centered at these locations also
complicates matters \citep{rn95}.

     Finally, we also experiment with various two-component fits.  These
fits all assume a weaker second component redward of the primary component,
as suggested by the Si~II fit in Figure~2, but we experiment with different
velocity separations and column density ratios for the two components.  The
column density ratio is typically assumed to be the same for all lines of
all atomic species in order to limit the number of free parameters.  We also
typically assume that the $T$ and $\xi$ values are identical for both
components.  We do not think that the FUSE or GHRS data have sufficient
resolution to accurately derive T and $\xi$ at all, let alone for both
components separately.  Nevertheless, we experimented with fits allowing T
and $\xi$ to be different for the two components and our total column density
measurements do not change significantly.  The fit to the data shown in
Figure~4 is an example of one of the global two-component fits, although only
the six strongest Lyman lines are shown in the figure.  This particular fit
assumes a velocity separation of 17 km~s$^{-1}$ and a very large column
density ratio of 250:1 for the two components, explaining why the second
component (dashed lines) is only seen for the H~I lines and not the
optically thin D~I lines.

     As described above, a two-Gaussian LSF is a free parameter of our fits.
In Figure~5, we show examples of LSFs that were derived for various fits, not
only to the WD~1634-573 data but also to FUSE observations of another white
dwarf, WD~2211-495, which are analyzed in detail by \citet{gh01}.
The solid lines are LSFs derived from fits to the LWRS data, and the
dotted lines are LSFs from the MDRS data.  There is significant scatter in
the estimated LSF shape, which indicates the level of uncertainty in this
average FUSE LSF, and it should also be pointed out once again that the real
FUSE LSF will be wavelength dependent.  There is no obvious systematic
difference between the LWRS and MDRS profiles, and there are not any obvious
systematic differences between the LSFs derived from the WD~1634-573 and
WD~2211-495 data sets.  The dashed line in Figure~5 is the average of all
these LSFs, which can be approximated by the sum of two Gaussians with the
following parameters:  $\lambda_{1}=0.01$ pix, $FWHM_{1}=10.40$ pix, and
$A_{1}=0.879$ for the first Gaussian, and $\lambda_{2}=0.82$ pix,
$FWHM_{2}=25.26$ pix, and $A_{2}=0.117$ for the second ($\lambda$=centroid
position, $FWHM$=full width at half-maximum, and $A$=amplitude).  The width
of this average LSF suggests a spectral resolution of
$R\equiv \lambda/\Delta\lambda \approx 15,000$.

     The FUSE LSF has broad wings, which is the reason the flux in the
saturated core of the broad H~I Lyman absorption lines does not fall to zero
(see Figs.\ 1, 3, and 4).  The prominence of the line wings of our derived
LSFs is determined almost entirely from the Lyman lines.  When we fit the
data without the H~I lines (see above), the broad wings disappear since the
remaining lower opacity line profiles are not nearly as sensitive to the
far wings of the LSF.

     In reality, the FUSE LSF will vary with wavelength, so we tried fits
with the LSF allowed to vary within the spectrum.  These trials did not yield
fit parameters statistically different from those obtained assuming an
invariable LSF.  However, it should be pointed out that there is not enough
information within the data to accurately determine a wavelength dependent
LSF, so our attempt to do so may not have fully accounted for systematic
errors associated with the wavelength dependence of the true LSF.
Variability in the LSF could be at least partially responsible for the
discrepancies in the fits to the H~I lines in Figure~4.  \citet{jwk01}
present further, detailed discussion about the difficulties in estimating
the LSF for FUSE data.

     All of the various fits that we try using different continua, different
data sets (i.e., LWRS or MDRS), and different types of multi-component fits
are visually inspected to make sure the spectra are fitted reasonably well.
After visibly poor quality fits are thrown out, we are left with a collection
of fits that we can use to determine the column densities and their
uncertainties.  For each column density, we have from these fits a set of
measured columns that lead to acceptable fits to the data.  This set defines
a range of acceptable values for that column density.  We choose the center
of that range as our best estimate for the column density and the half-width
of the range as the estimated uncertainty.

     For example, our set of $\log {\rm N(N~I)}$ values from various
acceptable fits is [14.51, 14.52, 14.59, 14.60, 14.61, 14.64, 14.66, 14.67,
14.71], where for the sake of brevity we have omitted repetition of identical
$\log {\rm N(N~I)}$ measurements.  From this set of numbers, we estimate
$\log {\rm N(N~I)}=14.61\pm 0.10$ since 14.61 is at the center of the range
of values and 0.10 is the half-width of the range.  Uncertainties estimated
in this manner are not statistical errors, but we consider them to be
something like 2$\sigma$ uncertainties in the sense of being roughly
95\% confidence intervals.  They certainly are more conservative than
``1$\sigma$ uncertainties'' since we do not throw out 32\% of our acceptable
fits before estimating the errors.  The temperatures we find with
our fits are in the $T=4800-7500$~K range and the nonthermal velocities are
in the range $\xi=0-5$ km~s$^{-1}$.  However, as noted above the presence
of multiple components complicates the interpretation of these quantities.

     As mentioned near the beginning of this section, we performed a
second independent analysis of the
data using a different technique.  This technique uses a spectral fitting
code called Owens.f, developed by one of us (M.\ L.) and the FUSE French
team.  A thorough description of the use of this code can be found
in \citet{gh01}, another of this series of papers on the first
deuterium measurements with FUSE, which describes the measurement of
deuterium toward WD~2211-495.  The Owens code performs global fits to many
absorption lines simultaneously by $\chi^{2}$ minimization.  All the data
from different FUSE channels and different observations are fitted
simultaneously, allowing for redundant coverage of a given spectral line.
The background continuum for each line, represented as a polynomial, is left
as a free parameter of the fit.  The width of the LSF is allowed to vary
independently for each line to try to account for expected LSF variations
with wavelength in the FUSE spectra.  Only the optically thin lines are
included in the fit in order to avoid systematic errors associated with the
structure of the line of sight and incorrect knowledge of the LSF.
Uncertainties (2$\sigma$) are estimated from an analysis of
$\Delta\chi^{2}$ contours \citep[see][]{gh01}.  The best fit
yielded a $\chi^{2}=3629.1$ with 3076 degrees of freedom.

     Despite the differences between the two fitting procedures, the results
are consistent, with overlapping error bars.  The final column density
measurements and 2$\sigma$ uncertainties listed in Table~4 represent a
synthesis of these two independent analyses, and are consistent with the
results of both.  We do not list a Si~II column density because of the
possible blend with N~III (see above).  The Si~II $\lambda$1260 line
observed by HST/GHRS (see Fig.~2) is saturated and therefore does not
provide a precise column density, but the S~II $\lambda$1259 line (see
Fig.~2) and a weaker S~II $\lambda$1253 line do provide a useful S~II column
density that we list in Table~4 for completeness.

\section{DISCUSSION}

     Because there are many D~I, N~I, and O~I lines of varying strengths,
we believe our column density measurements for these species are quite
accurate, with 2$\sigma$ uncertainties of only 0.05--0.07 dex (see Table~4).
The uncertainty roughly doubles when we have only one or two unsaturated
lines to work with, as is the case for P~II, S~II, Ar~I, and Fe~II.  When
we have only saturated lines to work with, the uncertainties are much higher.
Thus, the H~I, C~II, C~III, and N~II column densities have uncertainties in
the 0.3--0.7 dex range.

     This problem with saturated lines exists for H~I
despite there being many lines to work with, at least one
of which (H~I Ly$\beta$) is at least partly out of the flat part of the
curve of growth (see above).  Nevertheless, our analysis suggests that FUSE
data cannot be used to derive a precise H~I column density toward
WD~1634-573.  The situation will likely be better for higher column density
lines of sight with at least Ly$\beta$ and Ly$\gamma$ completely out of the
flat part of the curve of growth.  However, \citet{ml01} express concern
that undetected hot H~I components, such as the heliospheric and
astrospheric absorption previously observed toward nearby cool stars
\citep*{bew00}, could potentially affect any measurement of H~I
column densities from the Lyman absorption lines, making it even harder to
assess systematic errors for these analyses.

     Because our H~I measurement ($\log {\rm N(H~I)}=18.6\pm 0.4$) is
not very precise, we must look elsewhere for an H~I measurement
that we can use to compute an accurate D/H value for this line of sight.
\citet{jbh98} estimated $\log {\rm N(H~I)}\approx 18.8$
from their analysis of IUE observations of the Ly$\alpha$ line.  \citet{rn95}
found $\log {\rm N(H~I)}=19.0\pm 0.4$ from ORFEUS observations
of the same H~I Lyman lines that we have analyzed using FUSE.  Our results
are consistent with ORFEUS, but the error bars are large for both
measurements.  The most accurate measurement appears to be from EUVE
observations of H~I Lyman continuum absorption shortward of 912~\AA.  In a
simultaneous analysis of both ORFEUS and EUVE data, which allowed both
photospheric parameters (e.g., $T_{eff}$ and $\log g$) and the interstellar
H~I column density to vary, \citet{rn96} finds
${\rm N(H~I)}=(7\pm 2)\times 10^{18}$ cm$^{-2}$
(i.e., $\log {\rm N(H~I)}=18.85\pm 0.12$).  See also \citet{sj97}
for more details on the photospheric analysis.  Based on this H~I value
and our D~I measurement ($\log {\rm N(D~I)}=14.05\pm 0.05$), which are both
quoted with 2$\sigma$ uncertainties, we find that
${\rm D/H}=(1.6\pm 0.5)\times 10^{-5}$ toward WD~1634-573.  This value is
consistent with the LIC value of ${\rm D/H}=(1.5\pm 0.1)\times 10^{-5}$
\citep{jll98}.

     The Ar~I/H~I abundance ratio, $\log {\rm Ar~I/H~I}=-5.59\pm 0.17$, is
close to the B star Ar abundance of $\log {\rm Ar/H}=-5.50$ \citep{deh90}.
This is in contrast to the results of \citet{ebj00}, who
found significantly lower Ar~I/H~I ratios toward 3 stars
($\log {\rm Ar~I/H~I}\approx -5.9$), which they attributed to a higher
ionization state of Ar compared with H due to the high photoionization cross
section of Ar.  This provides important evidence that the LISM is in fact in
photoionization equilibrium.  Perhaps the WD~1634-573 line of sight is more
shielded from photoionization, allowing the ionization state of Ar to be
closer to that of H.  The apparently high N~II/N~I ratio (see
Table~4) might suggest otherwise, although uncertainty in the N~II column is
large.

     Due to charge exchange interactions, the ionization states of D, H, and
O are closely coupled.  This means that the total D/O gas phase abundance
ratio is well approximated by the D~I/O~I ratio.  Charge exchange is also
important between D, H, and N, although to a lesser extent, so
${\rm D/N}\approx {\rm D~I/N~I}$.  Our measurements of D~I, O~I, and N~I
in Table~4 yield ${\rm D~I/O~I}=0.035\pm 0.006$ and
${\rm D~I/N~I}=0.27\pm 0.05$ (2$\sigma$ errors).

     These ratios can be used to estimate D/H when we multiply them by
previously measured values for the gas phase O/H and N/H ratios in the ISM.
\citet{dmm98} used GHRS observations of the O~I $\lambda$1356 lines of 13
OB stars that are more distant ($d=100-1000$ pc) than WD~1634-573 to
estimate ${\rm O/H}=(3.43\pm 0.15)\times 10^{-4}$, with no evidence
for variation within the ISM.  Note that we have increased the published
O/H value by 7.5\% to account for a revised f value for O~I $\lambda$1356
suggested by \citet{dew99}.  Likewise, \citet{dmm97} find
${\rm N/H}=(7.5\pm 0.4)\times 10^{-5}$ toward a similar sample of 7 OB
stars, also with no evidence of variation.  Note that these measurements
of O/H and N/H only consider the column densities of O~I, N~I, H~I, and
H$_{2}$, and do not take into account the column densities of the ionized
species O~II, N~II, or H~II.  Since ${\rm O~I/O~II}={\rm H~I/H~II}$ is a
good approximation, the O/H value should be relatively free
from inaccuracies induced by ionization state issues, but the N/H value
is potentially susceptible to differences and variations in the relative
ionization states of N and H (see above).  Note also that the cited
uncertainties in O/H and N/H are 1$\sigma$ standard deviations of the mean
rather than simple standard deviations about the mean, meaning that the
scatter of the individual O/H and N/H measurements is actually larger than
suggested by the quoted uncertainties, potentially allowing for a larger
variation of O/H and N/H within the ISM than is suggested by those
uncertainties.

     In any case, multiplying our D~I/O~I measurement with the
\citet*{dmm98} mean O/H value results in
${\rm D/H}_{O}=(1.2\pm 0.2)\times 10^{-5}$, while multiplying our D~I/N~I
measurement with the \citet*{dmm97} mean N/H value results in
${\rm D/H}_{N}=(2.0\pm 0.4)\times 10^{-5}$ (2$\sigma$ errors).  These two
D/H values do not overlap, suggesting that the either the O~I/H~I and/or
N~I/H~I ratio toward WD~1634-573 is not consistent with the
\citet*{dmm97,dmm98} mean O/H and N/H values.  As suggested above, the
N~I/H~I ratio is more suspect due to the possibility of variations in the
relative ionization states of N and H within the LISM.
Furthermore, the N~I/H~I value measured toward G191-B2B is clearly lower
than the \citet*{dmm97} N/H value \citep{avm98,ml01}.  A more
extensive discussion of these issues is provided by \citet{hwm01}.
Note that the ${\rm D/H}_{O}$ measurement is a bit lower
than the accepted LIC value of ${\rm D/H}=(1.5\pm 0.1)\times 10^{-5}$.

\section{SUMMARY}

     We have used FUSE observations to measure column densities for many
atomic species toward the white dwarf WD~1634-573, which are listed in
Table~4.  These measurements and uncertainties represent a synthesis between
the results of two separate, independent analyses.  We feel the use of
independent analyses is the best possible method for assessing the magnitude
of systematic errors in this type of study.  Of particular significance
is the measurement of D~I, because of
the importance of the deuterium-to-hydrogen (D/H) ratio for cosmology and
studies of Galactic chemical evolution.  We find that a previously published
value for the H~I column density from EUVE is more precise than what we can
measure from FUSE, although our FUSE measurement is consistent with the EUVE
result.  Combining the FUSE D~I column density with the EUVE H~I column
density yields ${\rm D/H}=(1.6\pm 0.5)\times 10^{-5}$ (2$\sigma$
uncertainty), consistent with the LIC value of
${\rm D/H}=(1.5\pm 0.1)\times 10^{-5}$ \citep{jll98}.  Combining our
WD1634-573 measurement with others in this first series of FUSE D/H
measurements in the LISM, \citet{hwm01} find that
${\rm D/H}=(1.51\pm 0.07)\times 10^{-5}$ within the Local Bubble, which
extends $\sim 100$~pc from the Sun in most directions.  There is no
evidence for variation within the Local Bubble, but the dispersion of
Galactic D/H measurements increases significantly if IMAPS and
{\em Copernicus} measurements of longer lines of sight are considered.

     We can also estimate D/H for WD1634-573 by computing D~I/O~I and
D~I/N~I ratios from our FUSE measurements, and multiplying them by mean gas
phase O/H and N/H ratios derived for the ISM by \citet*{dmm97,dmm98}.
However, these yield inconsistent D/H values, suggesting that either
O~I/H~I and/or N~I/H~I toward WD~1634-573 is not consistent with the O/H
and N/H ratios from \citet*{dmm97,dmm98}.  The
N~I/H~I value is the more likely culprit since the ionization state of N is
more likely to vary relative to H than is the case for O, and the work of
\citet{avm98} and \citet{ml01} shows that N~I/H~I toward G191-B2B is
significantly lower than the \citet*{dmm97} N/H
value.  The D/H value computed from D~I/O~I and O/H,
${\rm D/H}_{O}=(1.2\pm 0.2)\times 10^{-5}$, is slightly lower than
the LIC value quoted above.  We note that the published O/H and
N/H values for the ISM are mean values derived from observations of stars
located much farther away than WD~1634-573 ($d=37.1\pm 2.6$~pc) and thus may
not be representative of gas in the nearby ISM.  The importance of FUSE
D~I/O~I and D~I/N~I measurements is discussed further by \citet{hwm01}.

\acknowledgments

     We would like to thank R.\ Napiwotzki for information about his
EUVE and ORFEUS studies of WD~1634-573, and we would like to thank J.\ C.\
Howk and D.\ York for useful comments on the manuscript.
This work is based on data obtained for the Guaranteed Time Team by the
NASA-CNES-CSA FUSE mission operated by the Johns Hopkins University.
Financial support to U.\ S.\ participants has been provided by NASA
contract NAS5-32985.  French participants are supported by CNES.  The first
author is supported in part by NASA grant NAG5-9041 to the University of
Colorado.

\clearpage

\begin{deluxetable}{lc}
\tablecaption{WD 1634-573 Properties}
\tablecolumns{2}
\tablewidth{0pt}
\tablehead{
  \colhead{Property} & \colhead{Value}}
\startdata
Other Name & HD 149499B \\
Spectral Type & DO \\
RA (2000) & 16:38:30 \\
DEC (2000) & $-57^{\circ}28.2^{\prime}$ \\
Gal.\ long.\ (deg) & $329.9$ \\
Gal.\ lat.\ (deg)& $-7.0$ \\
V & 11.7 \\
B--V & $-0.39$ \\
Distance (pc) & $37.1\pm 2.6$\tablenotemark{a} \\
$T_{eff}$ (K) & $49,500\pm 500$\tablenotemark{b} \\
$\log g$ & $7.97\pm 0.08$\tablenotemark{b} \\
\enddata
\tablenotetext{a}{Perryman et al.\ (1997).}
\tablenotetext{b}{Napiwotzki et al.\ (1995).}
\end{deluxetable}

\begin{deluxetable}{clccc}
\tablecaption{FUSE Observations}
\tablecolumns{4}
\tablewidth{0pt}
\tablehead{
  \colhead{Observation ID} & \colhead{Date} & \colhead{Aperture} &
    \colhead{\# of} & \colhead{Exp.\ Time} \\
  \colhead{} & \colhead{} & \colhead{} & \colhead{Exposures} & \colhead{(s)}}
\startdata
M1031101 & 2000 March 30 & LWRS &  8 &  5396 \\
M1031102 & 2000 April 12 & LWRS &  8 &  5448 \\
M1031103 & 2000 May 16   & LWRS & 12 &  8562 \\
S5140201 & 2000 April 13 & MDRS & 23 &11,464 \\
M1031104 & 2000 May 17   & MDRS & 14 &  6761 \\
\enddata
\end{deluxetable}

\begin{deluxetable}{lcccclccc}
\tabletypesize{\footnotesize}
\tablecaption{Detected ISM Lines}
\tablecolumns{10}
\tablewidth{0pt}
\tablehead{
  \colhead{Ion} & \colhead{$\lambda_{rest}$} & \colhead{f} &
    \colhead{W$_{\lambda}$ (m\AA)\tablenotemark{a}} & \colhead{} &
  \colhead{Ion} & \colhead{$\lambda_{rest}$} & \colhead{f} &
    \colhead{W$_{\lambda}$ (m\AA)\tablenotemark{a}}}
\startdata
H I & 913.480 & $1.29\times10^{-4}$ & \nodata         &
 & N I & 953.970 & $3.36\times10^{-2}$ & $31.6\pm 4.6$   \\
H I & 913.641 & $1.49\times10^{-4}$ & \nodata         &
 & N I & 954.104 & $4.43\times10^{-3}$ & $13.9\pm 5.0$   \\
H I & 913.826 & $1.70\times10^{-4}$ & \nodata         &
 & N I & 963.990 & $1.61\times10^{-2}$ & $25.5\pm 3.5$   \\
H I & 914.039 & $1.97\times10^{-4}$ & \nodata         &
 & N I & 964.626 & $1.04\times10^{-2}$ & $14.9\pm 8.5$   \\
H I & 914.286 & $2.30\times10^{-4}$ & $149.3\pm 12.4$ &
 & N I & 965.041 & $5.13\times10^{-3}$ & $14.2\pm 2.1$   \\
H I & 914.576 & $2.71\times10^{-4}$ & $154.7\pm 2.7$  &
 & N I & 1134.165 & $1.34\times10^{-2}$ & $32.0\pm 6.6$  \\
H I & 914.919 & $3.21\times10^{-4}$ & $158.3\pm 5.1$  &
 & N I & 1134.415 & $2.68\times10^{-2}$ & $38.1\pm 6.7$  \\
H I & 915.329 & $3.87\times10^{-4}$ & $157.7\pm 2.8$  &
 & N I & 1134.980 & $4.02\times10^{-2}$ & $43.1\pm 8.3$  \\
H I & 915.824 & $4.69\times10^{-4}$ & $156.1\pm 22.8$ &
 & N II & 915.613 & $1.45\times10^{-1}$ & $50.6\pm 14.7$ \\
H I & 916.429 & $5.78\times10^{-4}$ & $171.7\pm 6.5$  &
 & N II & 1083.991 & $1.03\times10^{-1}$ & $63.3\pm 5.0$ \\
H I & 917.181 & $7.23\times10^{-4}$ & $171.0\pm 9.1$  &
 & O I & 916.815 & $4.05\times10^{-4}$ & $6.6\pm 1.8$    \\
H I & 918.129 & $9.21\times10^{-4}$ & $184.7\pm 10.2$ &
 & O I & 919.658 & $8.13\times10^{-4}$ & $14.0\pm 5.2$   \\
H I & 919.351 & $1.21\times10^{-3}$ & $182.1\pm 6.2$  &
 & O I & 921.860 & $1.12\times10^{-3}$ & $14.3\pm 3.5$   \\
H I & 920.963 & $1.60\times10^{-3}$ & $200.1\pm 2.6$  &
 & O I & 924.952 & $1.59\times10^{-3}$ & $22.8\pm 1.7$   \\
H I & 923.150 & $2.22\times10^{-3}$ & $189.7\pm 17.8$ &
 & O I & 925.442 & $3.54\times10^{-4}$ & $5.0\pm 3.0$    \\
H I & 926.226 & $3.18\times10^{-3}$ & $204.9\pm 6.5$  &
 & O I & 929.517 & $2.36\times10^{-3}$ & $27.2\pm 5.3$   \\
H I & 930.748 & $4.82\times10^{-3}$ & $217.5\pm 4.5$  &
 & O I & 930.257 & $5.37\times10^{-4}$ & $8.2\pm 3.0$    \\
H I & 937.804 & $7.80\times10^{-3}$ & $224.6\pm 6.1$  &
 & O I & 936.630 & $3.73\times10^{-3}$ & $36.2\pm 3.0$   \\
H I & 949.743 & $1.39\times10^{-2}$ & $236.6\pm 7.8$  &
 & O I & 948.685 & $6.45\times10^{-3}$ & $39.7\pm 4.8$   \\
H I & 972.537 & $2.90\times10^{-2}$ & $257.5\pm 13.7$ &
 & O I & 950.885 & $1.57\times10^{-3}$ & $19.2\pm 5.4$   \\
H I & 1025.722 & $7.91\times10^{-2}$ & $378.6\pm 21.2$ &
 & O I & 971.738 & $1.48\times10^{-2}$ & $50.9\pm 3.7$   \\
D I & 937.548 & $7.81\times10^{-3}$ & $2.5\pm 2.0$    &
 & O I & 976.448 & $3.30\times10^{-3}$ & $34.1\pm 4.0$   \\
D I & 949.485 & $1.40\times10^{-2}$ & $7.2\pm 3.2$    &
 & O I & 988.578 & $5.42\times10^{-4}$ & \nodata         \\
D I & 972.272 & $2.90\times10^{-2}$ & $18.1\pm 4.8$   &
 & O I & 988.655 & $7.71\times10^{-3}$ & \nodata         \\
D I & 1025.443 & $7.91\times10^{-2}$ & $33.9\pm 4.2$  &
 & O I & 988.773 & $4.32\times10^{-2}$ & \nodata         \\
C II & 1036.337 & $1.23\times10^{-1}$ & $83.5\pm 15.7$ &
 & O I & 1039.230 & $9.20\times10^{-3}$ & $46.3\pm 8.3$  \\
C III & 977.020 & $7.62\times10^{-1}$ & $103.7\pm 9.6$ &
 & Si II? & 989.873 & $1.33\times10^{-1}$ & $36.5\pm 7.9$ \\
N I & 952.303 & $1.76\times10^{-3}$ & $3.0\pm 2.0$    &
 & P II & 963.801 & $1.46\times10^{0}$ & $10.0\pm 3.0$   \\
N I & 952.415 & $1.56\times10^{-3}$ & $2.0\pm 1.5$    &
 & Ar I & 1048.220 & $2.44\times10^{-1}$ & $26.4\pm 1.5$ \\
N I & 953.415 & $1.11\times10^{-2}$ & $25.8\pm 9.9$   &
 & Ar I & 1066.660 & $6.65\times10^{-2}$ & $8.9\pm 2.9$  \\
N I & 953.655 & $2.13\times10^{-2}$ & $25.1\pm 5.1$   &
 & Fe II & 1144.938 & $1.06\times10^{-1}$ & $11.8\pm 2.2$ \\
\enddata
\tablenotetext{a}{Uncertainties are 2$\sigma$.}
\end{deluxetable}

\begin{deluxetable}{lcc}
\tablecaption{Column Densities}
\tablecolumns{3}
\tablewidth{0pt}
\tablehead{
  \colhead{Species} & \# of Detected &
  \colhead{Log(Column Density)\tablenotemark{a}} \\
  \colhead{} & Lines & \colhead{(cm$^{-2}$)}}
\startdata
H I  & 21 & $18.6\pm 0.4$\tablenotemark{b}   \\
D I  &  4 & $14.05\pm 0.05$ \\
C II &  1 & $15.9\pm 0.7$   \\
C III&  1 & $15.4\pm 0.7$   \\
N I  & 10 & $14.62\pm 0.07$ \\
N II &  2 & $14.90\pm 0.35$ \\
O I  & 16 & $15.51\pm 0.06$ \\
P II &  1 & $12.08\pm 0.13$ \\
S II &  2 & $14.02\pm 0.09$\tablenotemark{c} \\
Ar I &  2 & $13.26\pm 0.12$ \\
Fe II&  1 & $13.08\pm 0.13$ \\
\enddata
\tablenotetext{a}{2$\sigma$ uncertainties.}
\tablenotetext{b}{Consistent with a more precise EUVE value of $18.85\pm 0.12$
  (Napiwotzki et al.\ 1996).}
\tablenotetext{c}{Measured from GHRS data.}
\end{deluxetable}

\clearpage

\begin{figure}
\plotfiddle{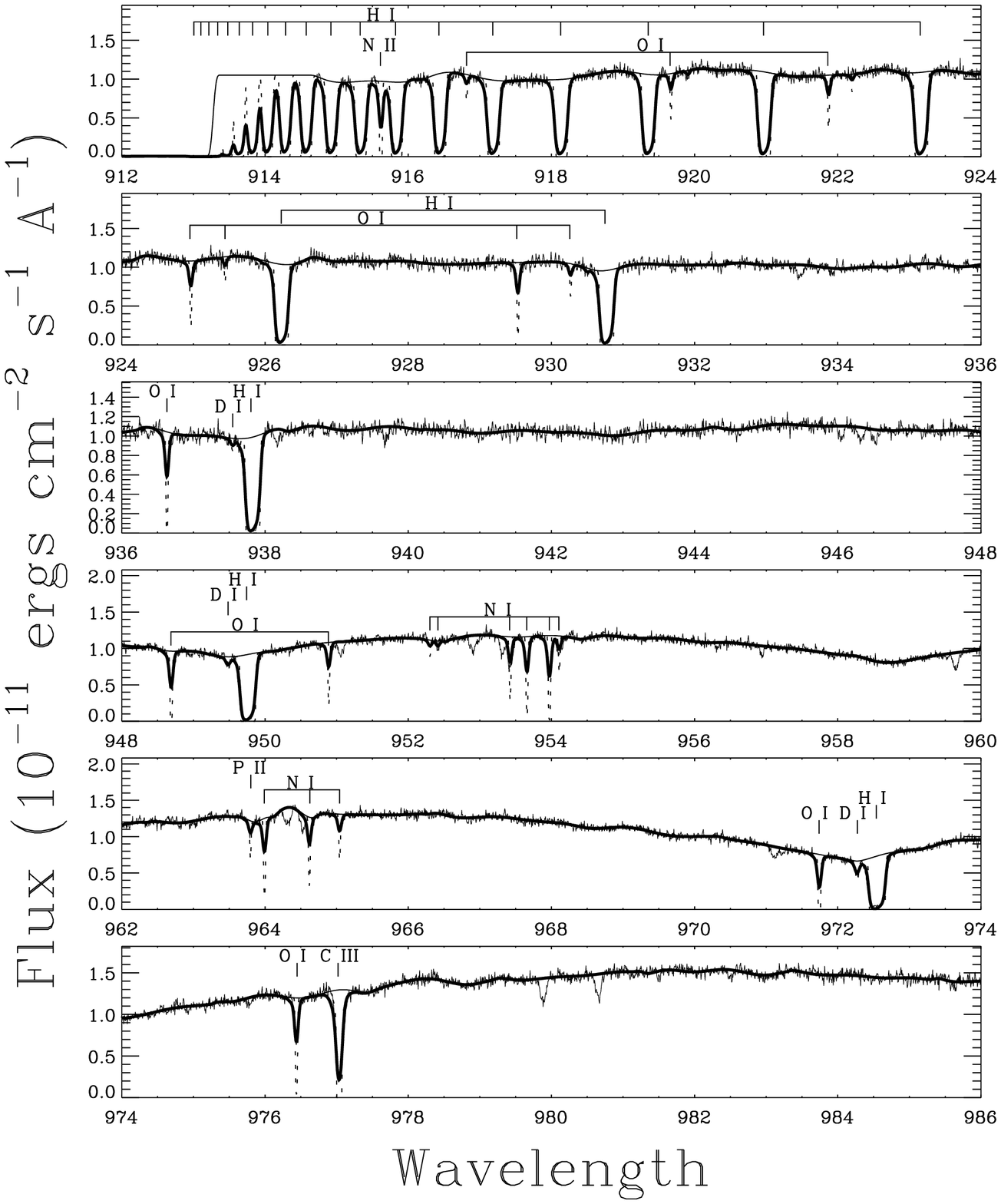}{6.5in}{0}{80}{80}{-340}{10}
\caption{Selected regions of the LWRS WD~1634-573 spectrum observed by FUSE
  containing various ISM absorption features, although the Ly$\beta$ line
  is actually MDRS data (see text).  Also shown is a global single
  component fit to the absorption lines, where the dotted and thick solid
  lines are the fit before and after convolution with the instrumental
  profile, respectively.  The instrumental profile is a free parameter
  of the fit.}
\end{figure}

\clearpage

\setcounter{figure}{0}
\begin{figure}
\plotfiddle{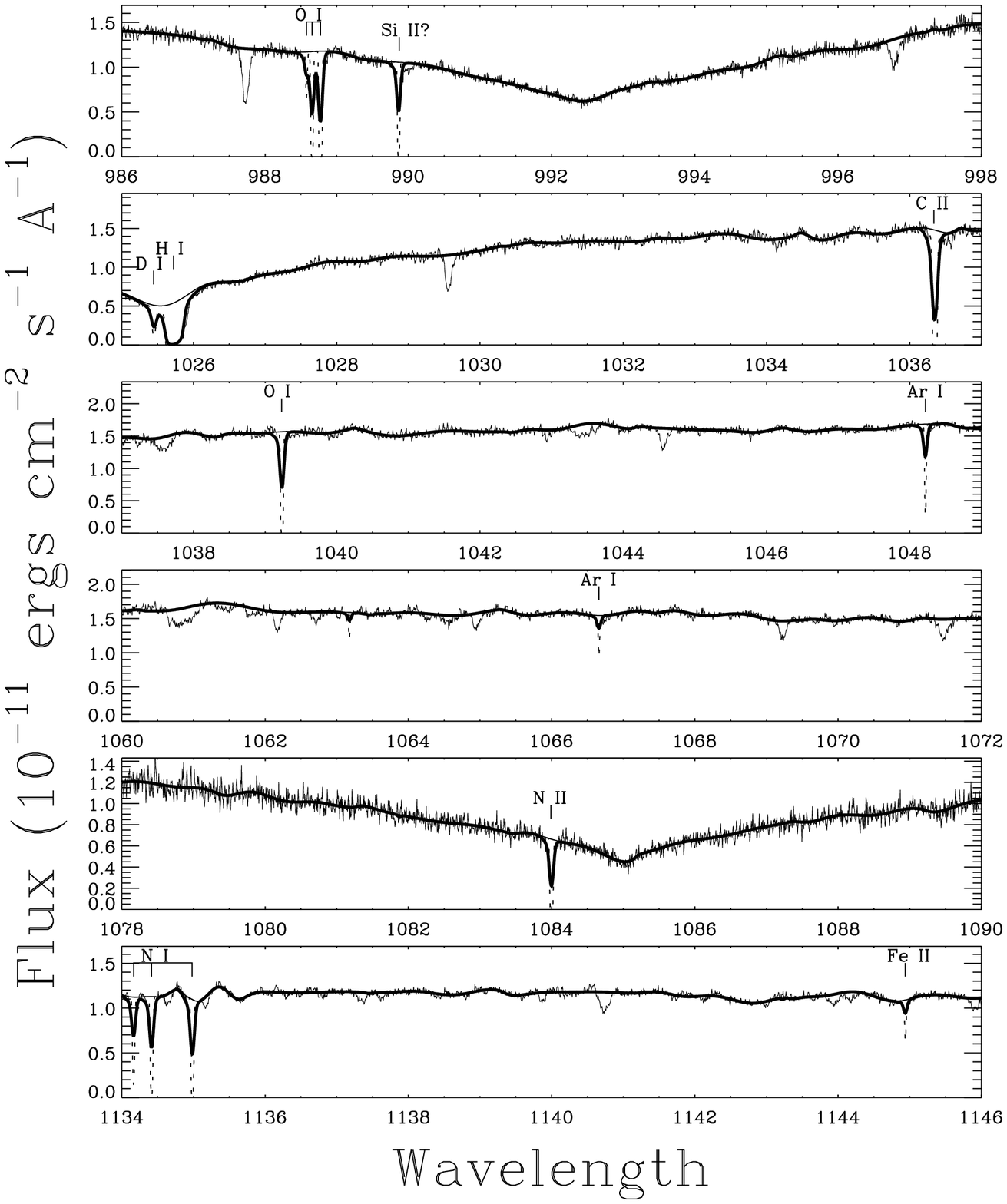}{6.5in}{0}{80}{80}{-340}{10}
\caption{(continued)}
\end{figure}

\clearpage

\begin{figure}
\plotfiddle{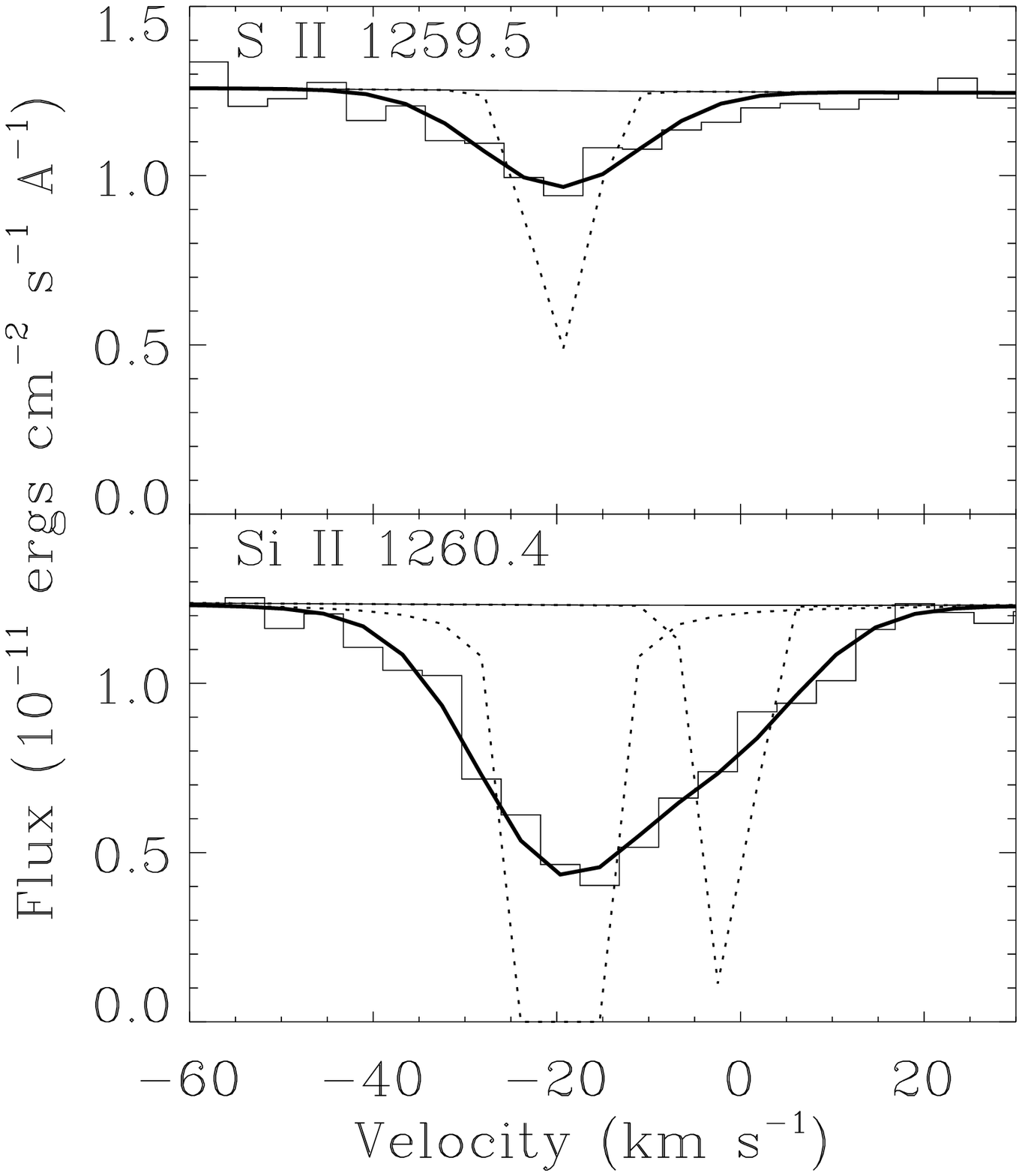}{4.5in}{0}{80}{80}{-270}{0}
\caption{HST/GHRS moderate resolution spectra of the interstellar
  S~II $\lambda$1259.519 and Si~II $\lambda$1260.422 absorption lines
  observed toward WD~1634-573, plotted on a heliocentric velocity scale.  The
  S~II line is fitted with a single absorption component, where the dotted
  and thick solid lines are the fit before and after convolution with the
  instrumental profile, respectively.  The opaque Si~II line is fitted with
  two absorption components, with one forced to be at the same velocity as
  the single component seen for S~II.}
\end{figure}

\clearpage

\begin{figure}
\plotfiddle{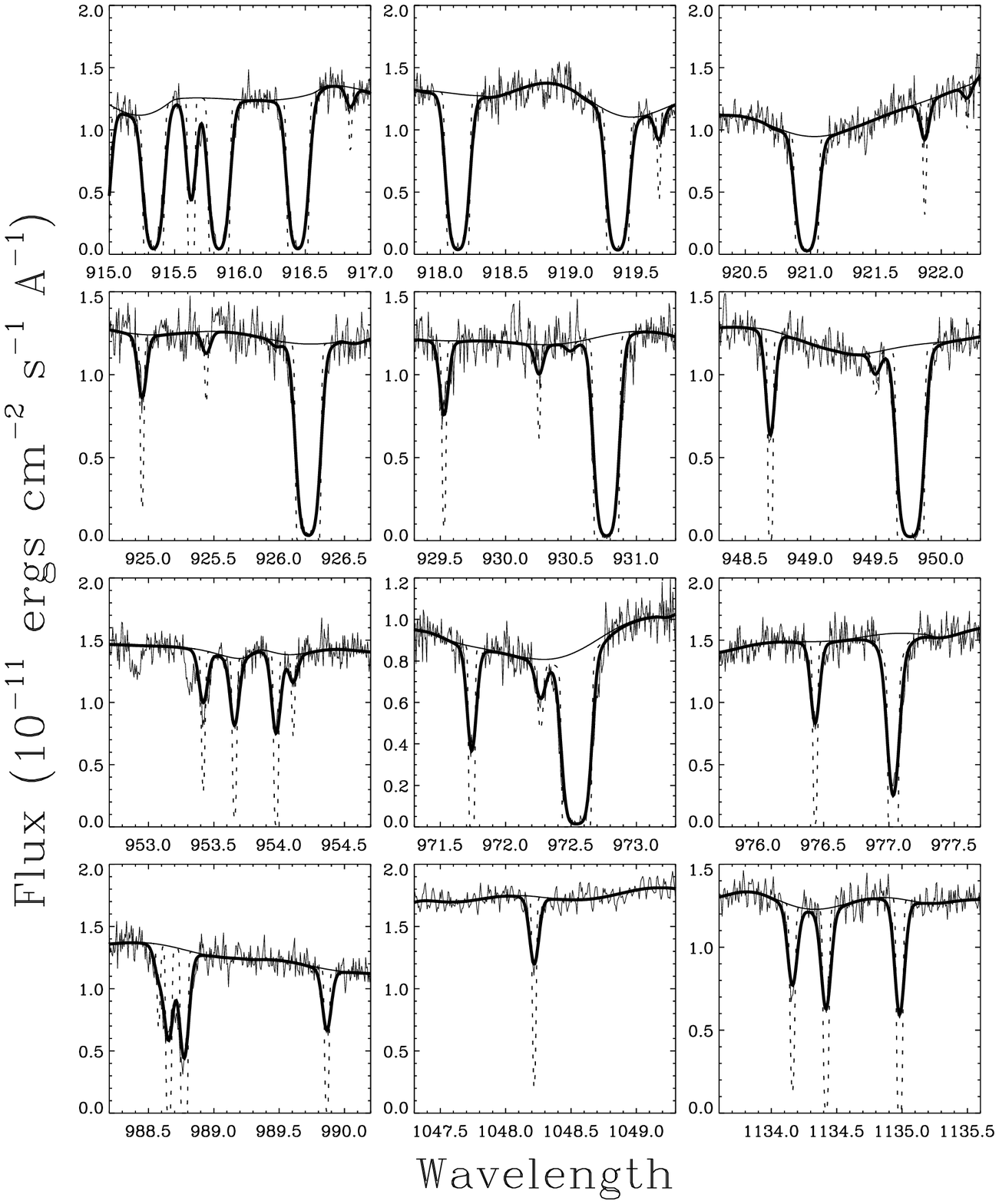}{6.5in}{0}{80}{80}{-340}{10}
\caption{Closeups of selected lines in the MDRS WD~1634-573 spectrum observed
  by FUSE.  See Fig.\ 1 or Table~2 for line identifications.  Also shown is a
  global single component fit to the absorption lines, where the dotted and
  thick solid lines are the fit before and after convolution with the
  instrumental profile, respectively.}
\end{figure}

\clearpage

\begin{figure}
\plotfiddle{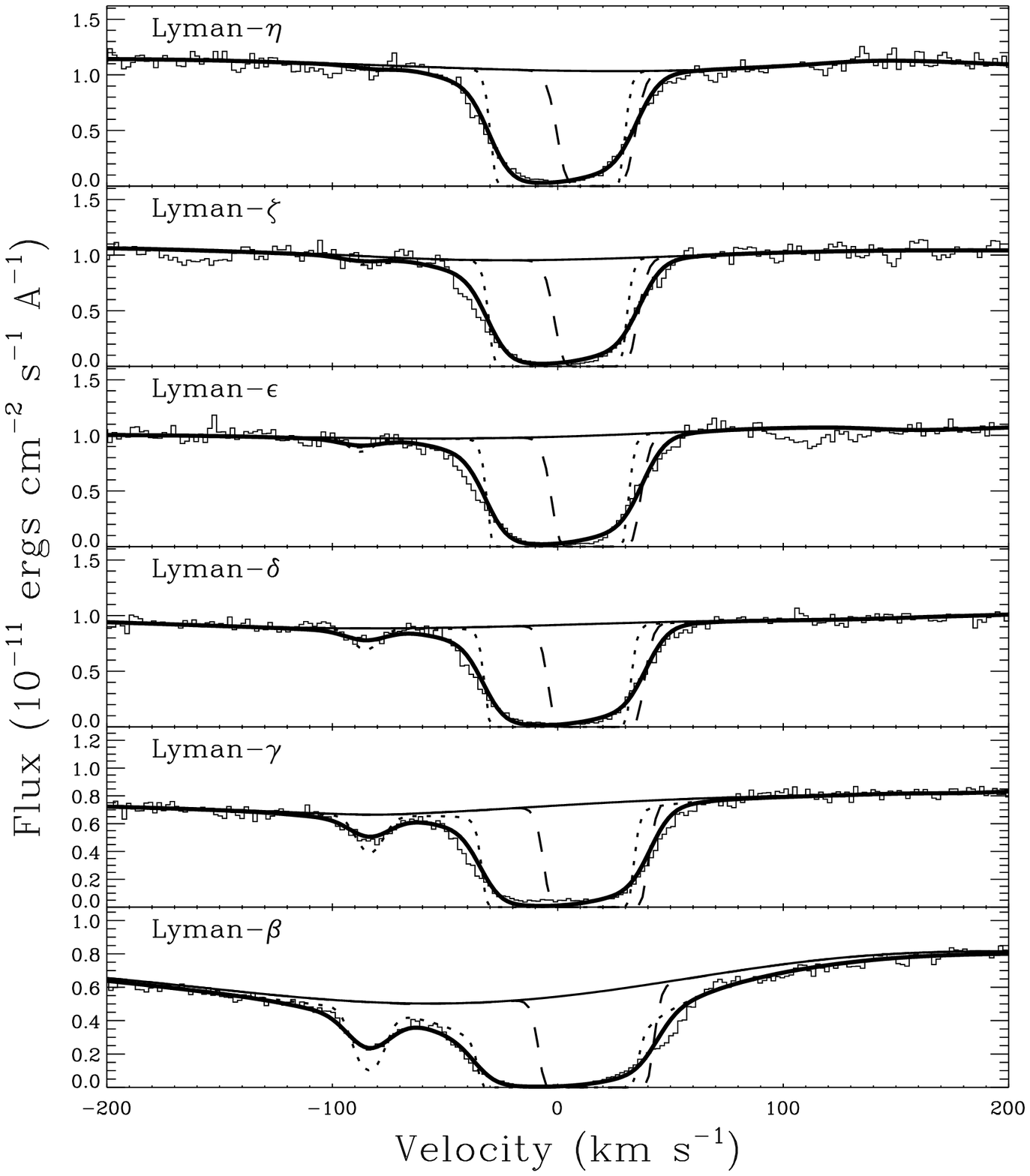}{6.5in}{0}{80}{80}{-340}{10}
\caption{The H~I and D~I Lyman lines observed by FUSE using the LWRS
  aperture, except for Lyman-$\beta$, which is MDRS data.  A two-component
  fit to these lines is shown, which is actually part
  of a global fit to all the ISM lines.  The dotted and dashed lines are the
  individual components, while the thick solid line is the combination of
  the two components after convolution with the instrumental profile, which
  is a free parameter of the fit.}
\end{figure}

\clearpage

\begin{figure}
\plotfiddle{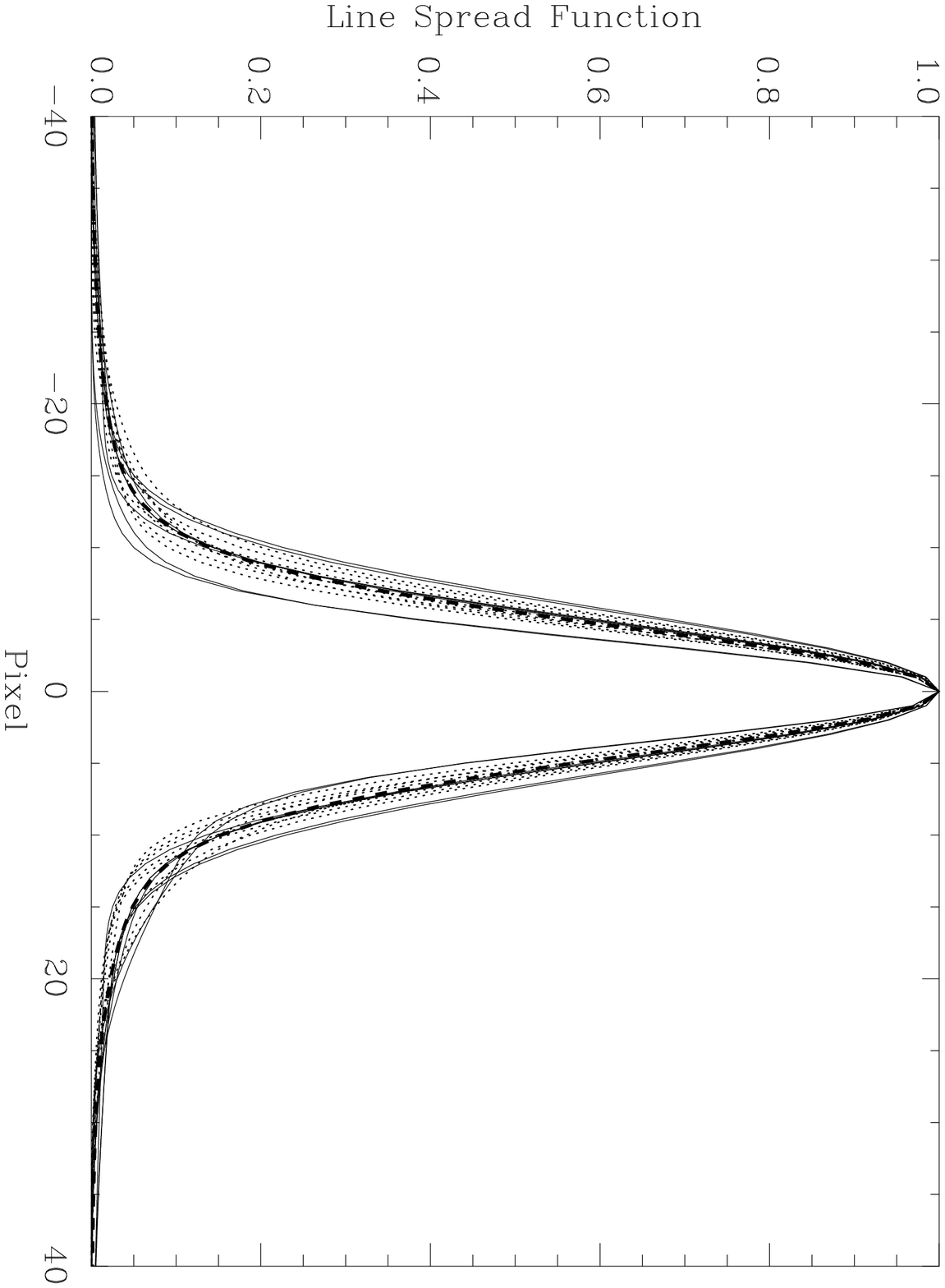}{3.5in}{90}{75}{75}{280}{0}
\caption{Various line spread functions derived from analyses of LWRS
  (solid lines) and MDRS (dotted lines) observations of WD~1634-573 and
  WD~2211-495.  The thick dashed line is the average LSF.  Two Gaussian
  fit parameters for this average LSF are provided in the text.}
\end{figure}

\end{document}